\documentclass{egpubl}
\usepackage{enumerate}
\SpecialIssueSubmission    
 \electronicVersion 

\ifpdf \usepackage[pdftex]{graphicx} \pdfcompresslevel=9
\else \usepackage[dvips]{graphicx} \fi

\PrintedOrElectronic

\usepackage{t1enc,dfadobe}
\usepackage{multicol,lipsum}
\usepackage{egweblnk}
\usepackage{cite}

\title[]%
      {Visual Subpopulation Discovery and Validation in Cohort Study Data}
 \author[S. Alemzadeh  \& T. Hielscher \& U. Niemann \& L. Cibulski \&T. Ittermann\& H. V\"olzke \& M. Spiliopoulou \& B. Preim]
       {S. Alemzadeh$^1$
        , T. Hielscher$^2$, U. Niemann$^1$, L. Cibulski$^1$,T. Ittermann$^3$, H. V\"olzke$^3$, M. Spiliopoulou$^2$, B. Preim$^1$
        \\
         $^1$Department of Simulation and Graphics, Otto-von-Guericke University Magdeburg, Germany
         \\
         $^2$Department of Technical and Business Information Systems, Otto-von-Guericke University Magdeburg, Germany
         \\$^3$University Medicine Greifswald, Germany
       }

\begin{document}


\maketitle

\begin{abstract}
Epidemiology aims at identifying subpopulations of cohort participants that share common characteristics (e.g. alcohol consumption) to explain risk factors of diseases in cohort study data. %
These data contain information about the participants' health status gathered from questionnaires, medical examinations, and image acquisition. %
Due to the growing volume and heterogeneity of epidemiological data, the discovery of meaningful subpopulations is challenging. %
Subspace clustering can be leveraged to find subpopulations in large and heterogeneous cohort study datasets. %
In our collaboration with epidemiologists, we realized their need for a tool to validate discovered subpopulations. %
For this purpose, identified subpopulations should be searched for independent cohorts to check whether the findings apply there as well. %
In this paper we describe our interactive Visual Analytics framework S-ADVIsED for SubpopulAtion Discovery and Validation In Epidemiological Data. %
S-ADVIsED enables epidemiologists to explore and validate findings derived from subspace clustering. %
We provide a coordinated multiple view system, which includes a summary view of all subpopulations, detail views, and statistical information. %
Furthermore, intervals for variables involved in a subspace cluster can be adjusted. This extension was suggested by epidemiologists. %
We investigated the replication of a selected subpopulation with multiple variables in another population by considering different measurements. %
As a specific result, we observed that study participants exhibiting high liver fat accumulation deviate strongly from other subpopulations and from the total study population with respect to age, body mass index, thyroid volume and thyroid-stimulating hormone.%
\end{abstract}
\section{Introduction}\label{introduction}
\begin{figure*}[!h]
  \centering
  \includegraphics[width=1\linewidth]{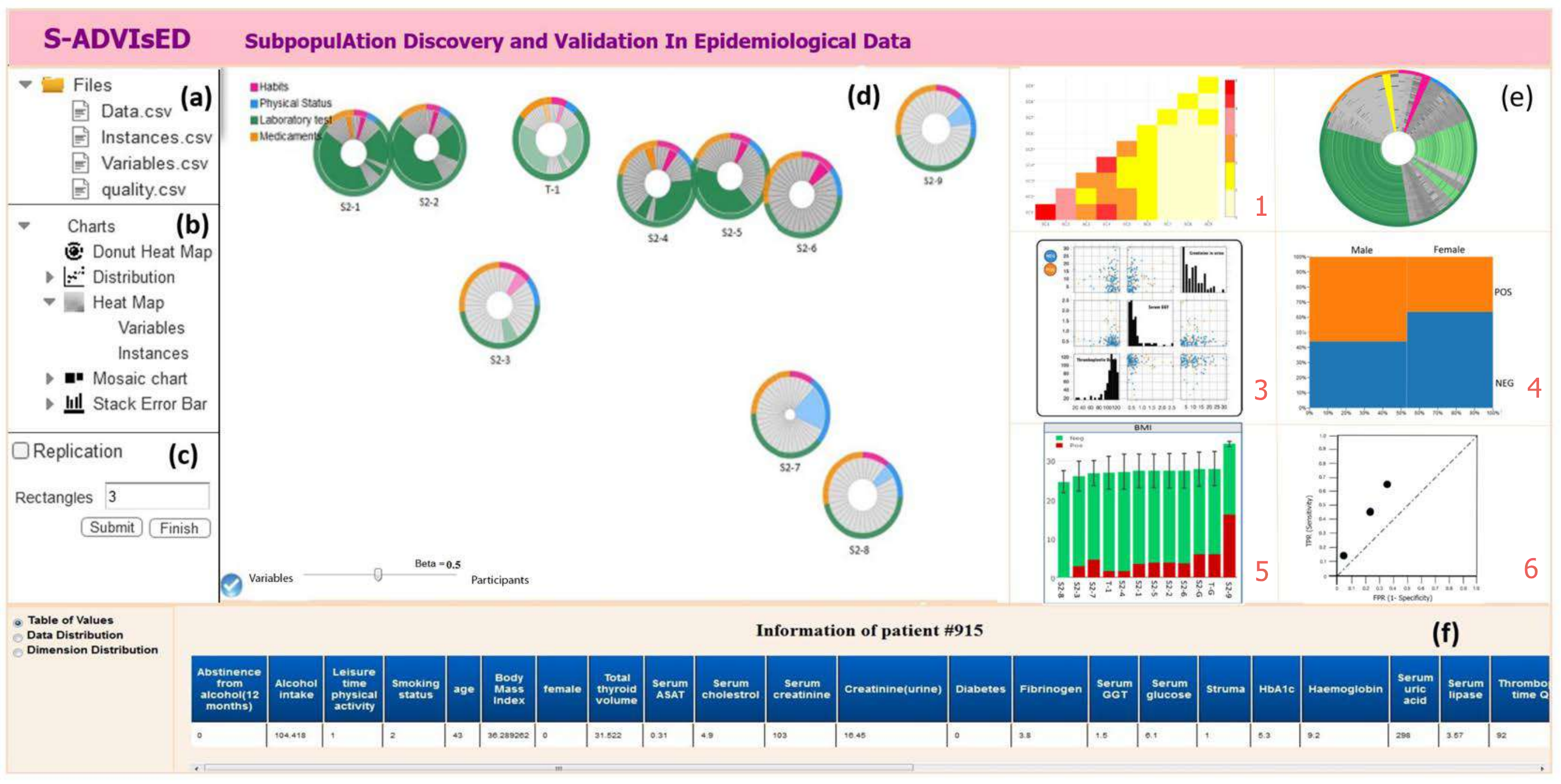}
  \parbox[t]{1\columnwidth}{\relax
           }
           \caption{\label{fig:screenshot}User interface of S-ADVIsED: (a) tree view of input files, (b) charts panel juxtaposes pairwise variable distributions of the selected subspace cluster by scatterplots (both numeric), mosaic charts (both categorical), and stacked errorbars (mixed numerical/ categorical) (c) replication settings, (d) global view of subspace clusters, (e) in-depth analysis, (f) statistical information.}
\end{figure*}
Epidemiologists investigate the factors which contribute to the outbreak of diseases as a prerequisite for preventive measures to counteract unfavorable health conditions. Thus, they identify risk factors related to life style, genetic predisposition, socio-demographic factors, and environmental factors as well as protective factors that reduce the likelihood of getting a disease \cite{woodward2013epidemiology}. This includes the analysis of interaction effects where two or more factors are considered simultaneously. Frequently, the combined influence of risk factors, such as smoking and above-average alcohol consumption, is stronger or weaker than the product of the individual factors. To do such investigations, they perform cohort studies, where they collect large quantities of data in a very standardized manner to get a comprehensive picture of the participants. These examinations are repeated several times to understand long-term effects and better assess causality. With the increasing amount of data, the traditional hypothesis-driven and statistics-focused approach cannot identify all possible correlations \cite{obenshain2004application}. %
In particular, it is usually impossible to identify subpopulations defined by several attributes that have a risk for a disease which strongly deviates from the global mean.
Data mining methods, in particular subspace clustering and subgroup discovery, are mature methods for the identification of subpopulations who share determinant factors. 
Subgroup discovery is a supervised learning task that aims at discovering interesting relations between sets of participants in a dataset with respect to the outcome. Identified patterns are expressed in the form of interpretable rules. For instance, a significant subpopulation could be phrased as ``While in the study population only 18\,\% exhibit goiter, in the subpopulation described by $BMI > 30.5\,kg/m^2 \wedge TSH \leq 1.5\,mU/l$ it is 52\,\%.'' %
Each condition in the rule antecedent corresponds to an axis-parallel hyperplane in the attribute space. %
Epidemiological data usually comprises also non-linear relationships between risk factor exposition and outcome (e.g. U-shaped or J-shaped relationships) \cite{preim2016visual}. %
Therefore, we prefer density-based subspace clustering since it allows for the discovery of arbitrary-shaped subspaces. %
Subspace clustering seeks for clusters in any subset of dimensions. %
However, due to their possibly complex shapes, they cannot be described in comprehensible rules and thus, they are less accessible for epidemiologists. Therefore, subspace clusters need to be transformed to hyper-rectangles such that they can be described as rules. In the assessment of clustering results and the transformation of subpopulations, visual representations are essential to enable the user to steer the process in detail and develop trust in the results. The latter is essential, since epidemiologists in general are skeptical to data mining results that may produce a very large amount of unreliable findings. To increase trust in the findings, epidemiologists start to share their data (which is very challenging due to data protection regulations). In our collaboration with epidemiologists we noticed their need for replication and validation of data mining findings.%
Our proposed S-ADVIsED framework combines visualization techniques and data mining concepts for discovery and validation of subspace clusters. S-ADVIsED allows to interactively explore subpopulations based on the user preferences. Our contributions in this paper include:
\begin{itemize}
    \item{Visual support for the identification of subpopulations in cohort study data}
    \item{Validating the findings in a second, independent cohort}
    \item{Exploration and comparison of subpopulations}
\end{itemize}
This paper extends a workshop contribution~\cite{AlemzadehEtAl:EuroVA2017} and describes S-ADVIsED in more detail. First we discuss the epidemiological background in Sec.~\ref{sec:EpidemiologyStudies}. We go on with related work in Sec.~\ref{sec:relwork}. In Sec.~\ref{sec:framework} we describe S-ADVIsED following, describing a use case scenario in Sec.~\ref{sec:usecase}. We discuss results in Sec.~\ref{sec:results} and conclude with a summary and future work in Sec.~\ref{sec:conclusion}.
\section{Epidemiological Background}
\label{sec:EpidemiologyStudies}
In this section we describe cohort study data, terms and methods used in this paper to address epidemiologists' requirements.
\subsection{Cohort Studies and Epidemiological Data}
Epidemiology research focuses on the determinants of health (risk and protective factors) and distributions in a specific population\cite{woodward2013epidemiology}. 
Distributions usually refer to the features which are related to the current situation of the population, including life style and socio-demographic status such as drinking alcohol, level of physical activity and marital status. Cohort studies are established to identify determinant factors in such subgroups w.r.t. a specific disease. Being aware of these factors helps to improve public health, e.g. by campaigning, safety measures and preventive examinations.\par
There are many sources of data in epidemiology. They contain information acquired by interview (e.g. sociodemographic, life style status, medication use), physical examination (e.g. measuring blood pressure and BMI), laboratory tests (e.g. diabetes and TSH) and medical images (e.g. MR images) that are used to measure imaging biomarkers and perform diagnosis. Combining such features consisting of categorical and continuous numeric attributes leads to a heterogeneous, high-dimensional and large data set. Preim et al.~\cite{preim2016visual} give an overview of large epidemiological studies. 
Our analyses are based on the Study of Health in Pomerania (SHIP) \cite{volzke2012study}. The study was performed in different waves, SHIP-0 (from 1997 to 2001), SHIP-1 (from 2002 to 2006), SHIP-2 (from 2008 to 2012) and SHIP-3 (from 2014 to 2016). Since this cohort gets older and smaller (due to dropouts from the study), a new cohort SHIP-TREND was established in parallel with the SHIP-2 study (between 2008 and 2012). This medium-sized study (initially 4308 participants) allows to study frequent diseases, such as diabetes and backpain. In this work we focused on the fatty liver as a widespread disorder. The information of the liver status we already extracted from medical images  by specialists. Participants with a liver fat concentration of more than 10\,\% are considered as positive for fatty liver and the rest as negative. 
In our samples, SHIP-2 participants are significantly older than TREND-0 participants, the proportion of females is higher and the distribution of the outcome differs slightly. To receive reliable quality estimates when validating one cohort's findings on another, we apply nearest neighbor propensity score matching \cite{Ho11} on age, sex and the outcome. After matching, each 694 of the SHIP-2 and TREND-0 participants remain, with age 55.5 $\pm$ 12.6 years, 46.5\,\% men, and 21.5\,\% fatty liver positive.
\label{sec:output}
Epidemiologists rely on a statistically-driven workflow: First, they formulate hypotheses based on observations made in clinical practice. %
Subsequently, they manually derive a list of possibly related variables and assess their correlation with the investigated outcome using statistical methods. %
For instance, regression modeling was successfully applied in a cohort study to show that air pollution exposure negatively affects birth outcomes \cite{Brauer:RegressionEpiExample08}. %
However, due to the continuously increasing dimensionality and heterogeneity of cohort study data, important associations might be overlooked. %
Furthermore, the goal of epidemiologists is not only to assess the global effect of a determinant of health, but also to find groups of study participants which are similar w.r.t. common protective and risk factors. %
\begin{figure}[h]
  \centering
  \includegraphics[width=1\columnwidth]{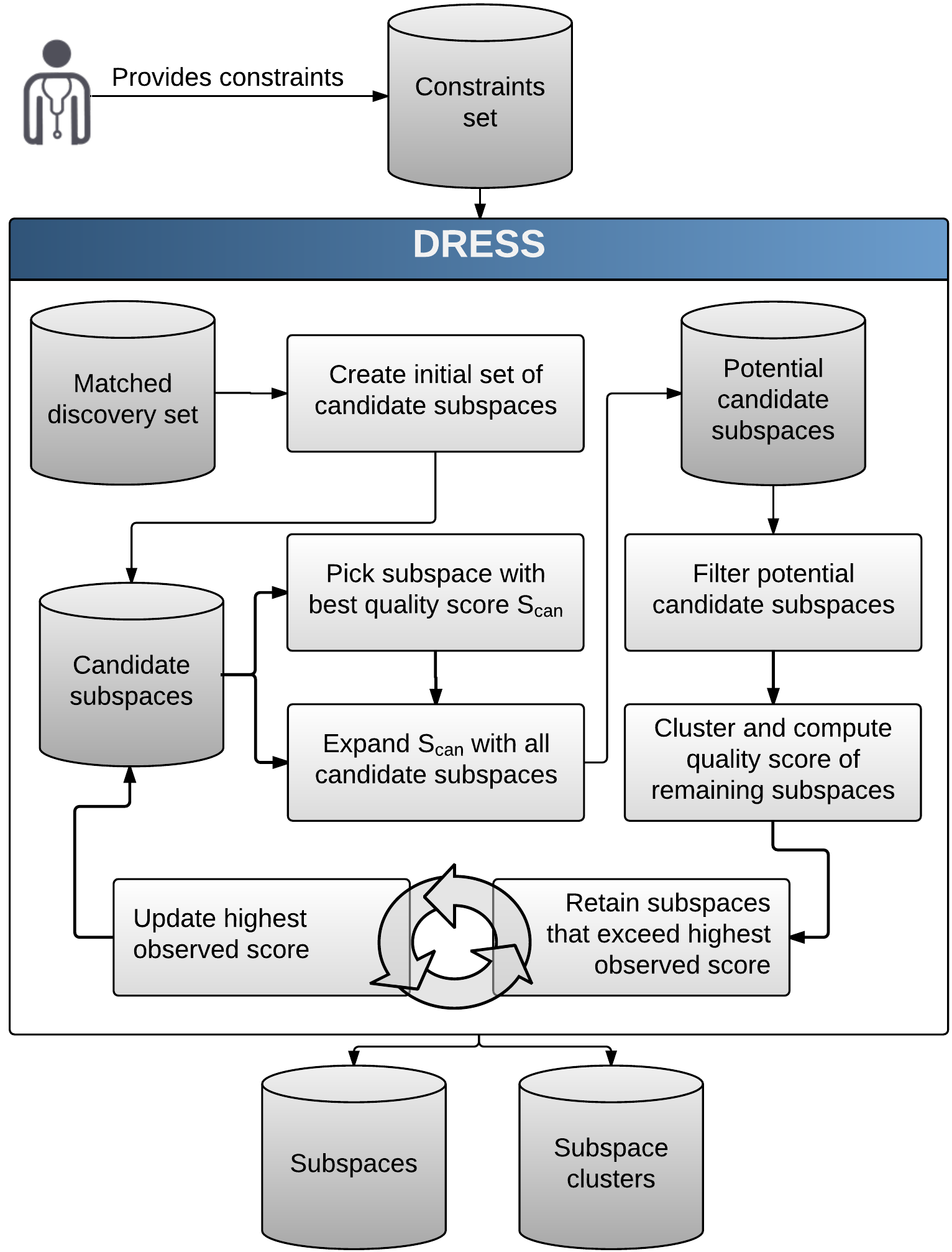}%
\caption{\emph{\label{fig:DRESS}General workflow of \emph{DRESS}. Constraints are given as input. Then, a forward selection commences to build better subspaces where must-link (not-link) constrained participants are more similar (dissimilar) and that contain clusters which satisfy more constraints. The algorithm delivers a ranking of subspaces and the contained subspace clusters.}}
\end{figure}
\subsection{Subspace Clustering}
Clustering methods are helpful for grouping similar individuals in cohort study data\cite{wheeler2007comparison}. %
However, their efficiency is hampered in high-dimensional feature space caused by an effect that is referred to as curse of dimensionality~\cite{parsons2004subspace}. Thus, these methods struggle to identify meaningful clusters when individuals are similar to each other for some variables (e.g. somatographic measurements), but not for others (e.g. age and sex). %
Subspace clustering algorithms can overcome this issue by automatically discovering interesting subpopulations in different subspaces. %
Subspace clustering involves two tasks~\cite{parsons2004subspace}: The selection of relevant subspaces and the actual clustering in these dimensions. %
While some methods integrate both tasks, others decouple them, e.g. search for appropriate (clusterable) subspaces first and apply a global clustering to the subspaces afterwards. The final result of subspace clustering is a set of clusters with a specific set of variables and individuals that are similar to each other w.r.t. the contributed variables. %
Niemann et al.~\cite{niemann2014subpopulation} investigate the performance of different unsupervised subspace clustering algorithms on a sample from SHIP. They reveal that algorithms are highly parameter-sensitive and non-intuitive to set, which is why epidemiologists are rather reluctant to use them in their routine. %
Furthermore, most subspace clustering algorithms are unsupervised, i.e. they do not consider the labels of the outcome. This leads to an enormous number of subpopulations regardless of whether associations with the outcome exist or not.
\emph{Constraint-based clustering.} Labeled participants are required to steer the algorithm in finding outcome-associated subpopulations. %
However, labeling often requires conducting expensive medical examinations, which makes the usage of fully supervised algorithms infeasible. %
Semi-supervised subspace clustering, like constraint-based subspace clustering, avoids these problems: they incorporate a small amount of background knowledge to find subspace clusters that best reflect the similarity between study participants w.r.t. the medical outcome. 
We employ the DRESS \cite{hielscher2016identifying} (\emph{D}iscovery of \emph{R}elevant \emph{E}xample-constrained \emph{S}ub\emph{S}paces) algorithm. %
The method is particularly useful for an application on cohort study data by incorporating expert knowledge and avoiding the necessity of class-labeled data and parameter optimization. %

Instead, \emph{DRESS} uses a small number of constraints on the similarity of study participants (must-link and not-link), which might be given by a medical expert to find clusters in subspaces that satisfy these constraints. To find subspace clusters that are highly associated with a disorder, such as fatty liver, an expert could define must-link constraints between participants that exhibit the same outcome, and not-link constraints between participants with and without fatty liver. The general workflow of \emph{DRESS} is as follows (Fig. \ref{fig:DRESS}):\\
\emph{DRESS} starts with a quality scoring of each subspace of cardinality one. Initially, these subspaces constitute the candidate set of subspaces. The subspace quality is scored by considering the distance between must-link and not-link constrained participants in the respective subspace as well as the proportion of satisfied constraints to all constraints. For the respective subspace, a must-link constraint is satisfied if both constrained participants lie within the same cluster and a not-link constraint is satisfied if both participants are members of different clusters.  \emph{DRESS} iteratively picks the best scored subspace $S_{can}$ and merges it with all remaining subspaces in the candidate set. To reduce complexity, the resulting subspaces are filtered by a reduced quality criterion (faster to compute than the full quality), i.e. if the calculated quality part is lower than for the original subspaces, the merged subspace is not further considered. For all subspaces that satisfy the filter criterion the full quality is calculated, which involves a density-based clustering with DBSCAN \cite{Ester96} where parameters are automatically determined \cite{Niemann15}. As soon as the quality of a subspace exceeds the highest yet observed quality $q_{best}$, \emph{DRESS} retains it as a candidate subspace for further extension, updates $q_{best}$ and stores all contained clusters. At the end of an iteration, $S_{can}$ and all merge candidates that led to a new $q_{best}$ are removed from the candidate set. \emph{DRESS} terminates when the candidate set is empty and returns a ranking of subspaces and their associated clusters.
Often, subspace clusters that have a high quality according to an \emph{interestingness measure} are very similar, e.g. they differ only in one dimension. To enable the analysis of a representative subset of subspace clusters, it is helpful to analyze such relations and show them graphically, e.g. as a hierarchy visualization \cite{Achtert:2007}.
\section{Related Work}
\label{sec:relwork}
This section provides an overview of related works for both the analysis  of cohort study data and the visualization of subspace clusters.%

\emph{Visual Analytics on Cohort Study Data.} 
Zhang et al. proposed an interactive visual analytics tool to analyze the cohort population\cite{zhang2014iterative}. Cohort Analysis via Visual Analytics (CAVA) comprises three main parts: The \emph{Cohort} part contains a set of people with similar properties, \emph{views} are a visual representation of the cohort and \emph{analytics} stand for the analyzing cohorts via interactive with the user. 
Krause et al. \cite{krause2016supporting} provided an interactive framework for Supporting Iterative Cohort Construction With Visual Temporal Queries (COQUITO).
The COQUITO tool enables the analyst to iteratively define subpopulations using visualizations. The analyst can add elements to the model by drag and drop (e.g. constraints). The analyst compares the results of constructed queries regarding their distributions and properties via treemaps and bar charts respectively.
Klemm et al. \cite{klemm2014interactive} presented a visual analytics system to identify subpopulations on the basis of data interactions using three global clustering on shape parameters characterizing the spinal canal to better understand backpain using SHIP data. In their results, all clustering algorithms were highly sensitive to parameters. Klemm et al. \cite{klemm20163d} enabled epidemiologists to enter regression formula and search for variable combinations related to an outcome, e.g. increased breast density. With heatmaps indicating strong correlations users are guided to potentially relevant factors. The analysis, however, does not involve any subpopulations.\\
Alemzadeh et al. \cite{AlemzadehEtAl:VCBM2017} presented an application for the visual analysis of missing data in longitudinal cohort study data. The presented framework enables the analyst to explore the missing values and check the predicted values after imputation process.
Niemann et al. \cite{NiemannEtAl:CBMS2017} proposed a hierarchy-based technique that clusters the rules produced by the subgroup discovery algorithm. Their main purpose is to summarize the rules produced by a subgroup discovery algorithm instead of manually filtering the rules. 

\emph{Visual Analysis of high dimensional data and Subspace Clusters.} Assent et al. presented a novel visualization technique for the exploration of the results of subspace clustering \cite{Assent:2007:VVS:1345448.1345451}. Their Visual Subspace Clustering Analysis (VISA) encoded each subspace cluster as a circle. The radius of circles depicts the size of subspace clusters. The color of each circle represents the dimensionality of subspace clusters.
The global overview of subspace clusters shows the similarity of subspace clusters w.r.t. their shared participants and dimensions. The more similar subspace clusters are closer to each other. They used multi-dimensional scaling (MDS)~\cite{wickelmaier2003introduction} to project the subspace clusters in a 2D layout. We also used this method to represent the similarity of subspace clusters.
VISA contains a matrix view to show in-depth information.
Subspace clusters with a higher number of objects may overlap the other subspace clusters and they become hidden from our view because circles have different sizes. 
Tatu et al. presented ranking measures for visualisation of scatterplots and parallel coordinates in order to ease and speed up exploration of high dimensional dataset~\cite{tatu2009combining}. In a further work they present ClustNail, which is a combination of several visualization methods to explore the subspace clustering result. They used circles to show subspace clusters \cite{tatu2012clustnails}. An ordering is performed to explain the similarity of different subspace clusters. ClustNail gives an importance to each dimension based on its variance. The size of the inner circle demonstrates the cluster size and the spike size indicates the importance of each dimension. To show in-depth information of subspace clusters, it provides a heat map of both involved and non-involved variables. Therefore, it will not lose the information of unselected variables. 
Achtert et al. presented DiSH as an algorithm for Detecting Subspace cluster Hierarchies~\cite{Achtert:2007}. DiSH is able to find clusters with different dimensionality, size, density and shape. Then, they visualize the relationship between hierarchy of subspace clusters by graphs.
Hund et al. in\cite{hund2016visual} proposed a multiple simultaneous view of information to explore large medical datasets. They used heat maps to represent the distances between subspace clusters. An aggregation table shows the values of cumulative cluster participants. Statistical information about data and object distribution is provided by bar charts. %

\section{Visual Analytics Support for Epidemiological Analysis}  
\label{sec:framework}
Epidemiologists mostly rely on statistical methods and simple visualizations. Data mining methods may be useful to them when they can trust their findings. A visual analytics system where data mining is not just a black box combined with explicit support for validation is essential to support epidemiologists. 
To generate hypotheses about subpopulations, subspace clustering is used to find groups of similar participants (recall Sec. \ref{sec:output}) regarding a specific disease. This requires a comparison of a subpopulation to either the overall population or another subpopulation.
A subspace clustering algorithm gives the vectors of variables and individuals that contribute to subspace clustering as an output. Exploration of subspace clusters is needed to get insight about patterns, distributions and assess the quality of discovered subpopulations.
Here, we propose S-ADVIsED as a visual analytics framework that combines several visualization methods for discovery, validation and comparison of subpopulations. S-ADVIsED is a web-based framework using d3.js library in JavaScript~\cite{bostock2011d3}. To model the global view of subspace clusters we have used dimple.js \footnote{{\url:http://dimplejs.org/}}  and for charts we employed plotly.js \footnote{{\url:http://plot.ly/javascript/}}libraries.

The screenshot of S-ADVIsED is shown in Fig.~\ref{fig:screenshot}. The interface is composed of six panels: 
\begin{itemize}
  \item{The files panel (a) displays input files including the dataset and subspace clustering result. }   
  \item{The charts panel (b) juxtaposes pairwise variable distributions of the selected subspace cluster by scatterplots (both numeric), mosaic charts (both categorical), and stacked errorbars (mixed numeric/categorical). }
  \item{The replication panel (c) contains information for replication process.}
  \item{The central panel (d) is a fixed view that provides a general overview of all subspace clusters and their characteristics. All interactions toward subspace clusters are carried out through this panel. }
  \item {The right side panel (e) includes analytics for subpopulation validation and charts based on the user-selected chart. In general, it presents detailed information of subspace clustering. }
  \item {The panel located in the footer provides statistical information about subspace clusters and participants. It enables users to get a simultaneous view to analyze subspace clusters.}

  
  \end{itemize}

\subsection{Exploration of Subspace Clusters}\label{sec:requirements}
 To get insight into the discovered subspace clusters, the relationship between involved variables, and to compare subpopulations, we need the visual analytics framework. Epidemiologists assess the interestingness of subspace clusters based on their intended measurements.
To evaluate discovered subpopulations and subpopulations\cite{hund2016visual2}, we have to consider the following requirements for the visualization of subspace clusters:

%

%

\begin{enumerate}[R1]
\item \label{dimensionality}\emph{Dimensionality:} 
Epidemiologists are more interested in low-dimensional subspace clusters.
Knowledge derived from subspace clustering should ultimately be transferred to clinical practice, i.e. contribute to prevention, diagnosis and treatment of diseases. %
Therefore, subspace clusters should be rather low-dimensional, since the addition of another variable might represent the necessity for the physician to conduct a further, possibly expensive medical examination. %
Also, low-dimensional subspace clusters follow the principle of scientific parsimony and are thus likely to be significant and not prone to overfitting. %
\\
\item \emph{Cluster Size:}\label{size} The number of participants in each subpopulation is an important factor. To support evidence of statistical significance, subspace clusters should cover at least 5\,\% of all study participants (in case of medium-sized studies as SHIP).
\\    
\item \emph{In-depth Information:}\label{depth} For each subspace cluster, a clear and compact overview visualization showing the distributions of both involved and non-involved variables should be displayed.
\\  
\item \emph{Cluster Compactness:}\label{compactness} Participants who belong to one subspace cluster should be similar to each other with respect to their involved dimensions. For example, when BMI is an involved dimension in one subspace cluster, then it is expected that all individuals are in the same interval, i.e. BMI between 20 and 24.
\\
\item \emph{Dimension Redundancy:}\label{dimredundancy} To compare different subspace clusters, it is necessary to consider the number of shared attributes to have a view of similarity between clusters.  
\\
\item \emph{Object Redundancy:}\label{insredundancy} It is necessary to indicate the overlap proportion of participants in different subpopulations. It helps epidemiologists to have an overview on the similarity between subpopulations in terms of shared participants.
\\     
\item \emph{Comparison with global mean:}\label{comparison} It is crucial to compare different subpopulations with the whole population. As an example, epidemiologists are interested in investigating subpopulations that differ strongly from other subpopulations w.r.t. outcome or a specified attribute. Actually, strong deviations represents subpopulations with a high relative risk.
\\
\item \emph{Dimension Variability:}\label{variability} 
Subspace clustering algorithms typically minimize the sparsity of data by ignoring variables with higher variance. It might be interesting for epidemiologists to pursue the reason for incorporating a high variance variable in a cluster. %
\end{enumerate}

\subsection{Validation of Findings}\label{sec:validation}
Subspace clusters may have arbitrary shape and subpopulations need to be defined as intervals in the form of hyper-rectangles (recall Sec. \ref{introduction}). Additionally, epidemiologists consider these findings not acceptable as long as they are not validated.\\
One strategy to validate the subspace clusters is replication. This means, if a specified subspace cluster can be reproduced in an independent population, then the discovered subspace cluster can be considered as a relevant subpopulation. In other words, the cluster description should be verified in another cohort. Therefore, it is necessary to let the epidemiologists to adjust the shape of the selected cluster in desired intervals and checks its reproducibility.
Different measures are specified by epidemiologists to check the reproducibility of the intended subpopulation \cite{cibulski2016visual}. 
\begin{enumerate}
\item Distribution: the distribution regarding to the involved variables and target variable in both replicated and original subpopulation should be similar.
\item Dimensionality: Both subpopulations should contain the same variables. 
\item Subpopulation size: The relative (5\% for example) number of participants in both clusters should be close to each other.
\item Deviation: Subpoulations are considered similar if they differ in the similar way from the global mean relative risk for a disorder regarding to the specific variables.
\end{enumerate}

In the following we describe the techniques we chose to fulfill the epidemiologists' requirements. To analyze the subspace clusters, we filtered the 25 subspace clusters with the highest p-value identified by DRESS. These high p-values indicate that the risk factor for fatty liver differs strongly from the global mean.
We also withdraw very similar clusters regarding their shared variables and subspace clusters. 

\begin{figure}[h]
  \centering{\includegraphics[width=.8\columnwidth]{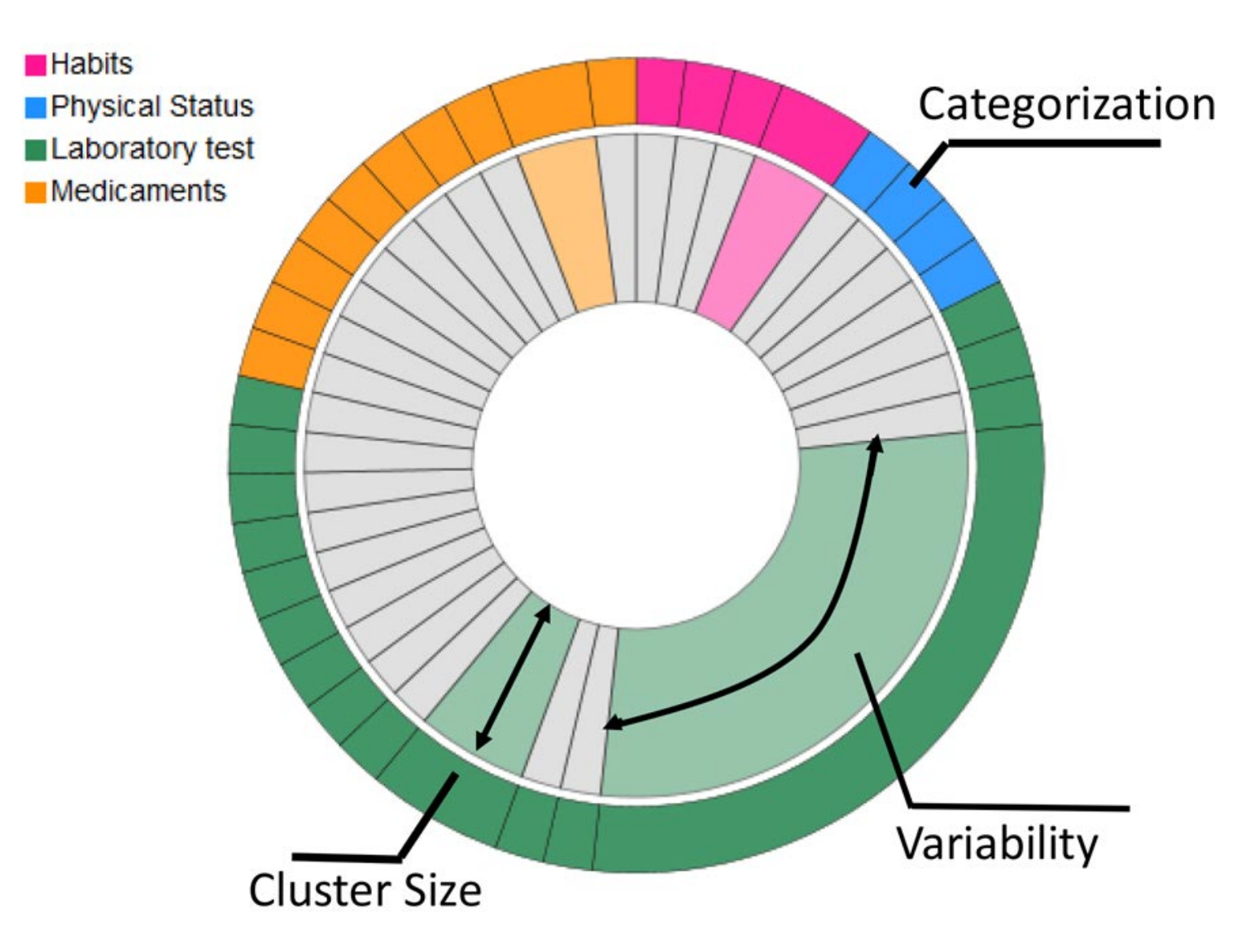}}%
\caption{\emph{\label{fig:donuts}Each subspace cluster is shown by a donut chart. Coloring is based on the categorization of variables and the radius of the donut depicts the cluster size.}}
\end{figure}

\subsubsection{Global Overview}\label{sec:globalview}
 In the Global view we have an overview of all subspace clusters and the main characteristics of subspace clusters:

We illustrate subspace clusters by donut charts, since they have a simple representation and we are able to encode enough information in them to show different specifications of subpopulations (Fig. \ref{fig:donuts}). The encoding characteristics are as follow:
    \begin{itemize}
         \item Sectors stand for variables and their size depicts their variability based on the variance (R\ref{variability}).
        \item The radius size of the donut chart depicts the cluster size. Bigger radius means it contains more participants (R\ref{size}). 
        
        \item The colored sectors represent involved variables in subspace clustering results and the grays are non-involved ones (R\ref{dimensionality}).
        
        \item Linking and brushing techniques are implemented to show variable overlaps. By clicking on each dimension the corresponding dimension in other subspace clusters will be highlighted.

    \end{itemize}
   
 Similar to Assent et al. \cite{Assent:2007:VVS:1345448.1345451}, we define the distance between subspace cluster $SC_{i}$ and $SC_{j}$ in subspaces  $S_{i}$ and $S_{j}$ as 
    \begin{equation} \label{eq:similarity}
\beta \left ( 1-\frac{\left | S_{i}\bigcap S_{j} \right | }{\left | S_{i}\bigcup S_{j} \right |} \right ) +
\ \left ( 1-\beta  \right ) \left ( 1-\frac{ \left | C_{i}\bigcap C_{j} \right |  }{ min\left \{ \left | C_{i} \right | , \left | C_{j} \right | \right \}   } \right )
\end{equation}
The first part of the equation reflects the fraction of shared variables between $S_{i}$ and $S_{j}$, whereas the second part quantifies the fraction of shared study participants. The expert can give more importance either to object overlap or subspace overlap by setting the weighting parameter $\beta$ to values below and above 0.5, respectively.

    To illustrate the distance of subspace clusters a MDS is employed to project the clusters in 2D space. MDS is frequently used for evaluating clustering\cite{Assent:2007:VVS:1345448.1345451,hund2016visual}.
    MDS gets a spatial representation of subspace clusters in a 2D space, indicating the similarity between resulting subspace clusters. A slider is employed to adjust the importance level of variable over participants overlaps for the visualization. 
     
 \begin{figure}[h]
  \centering
  {\includegraphics[width=1\columnwidth]{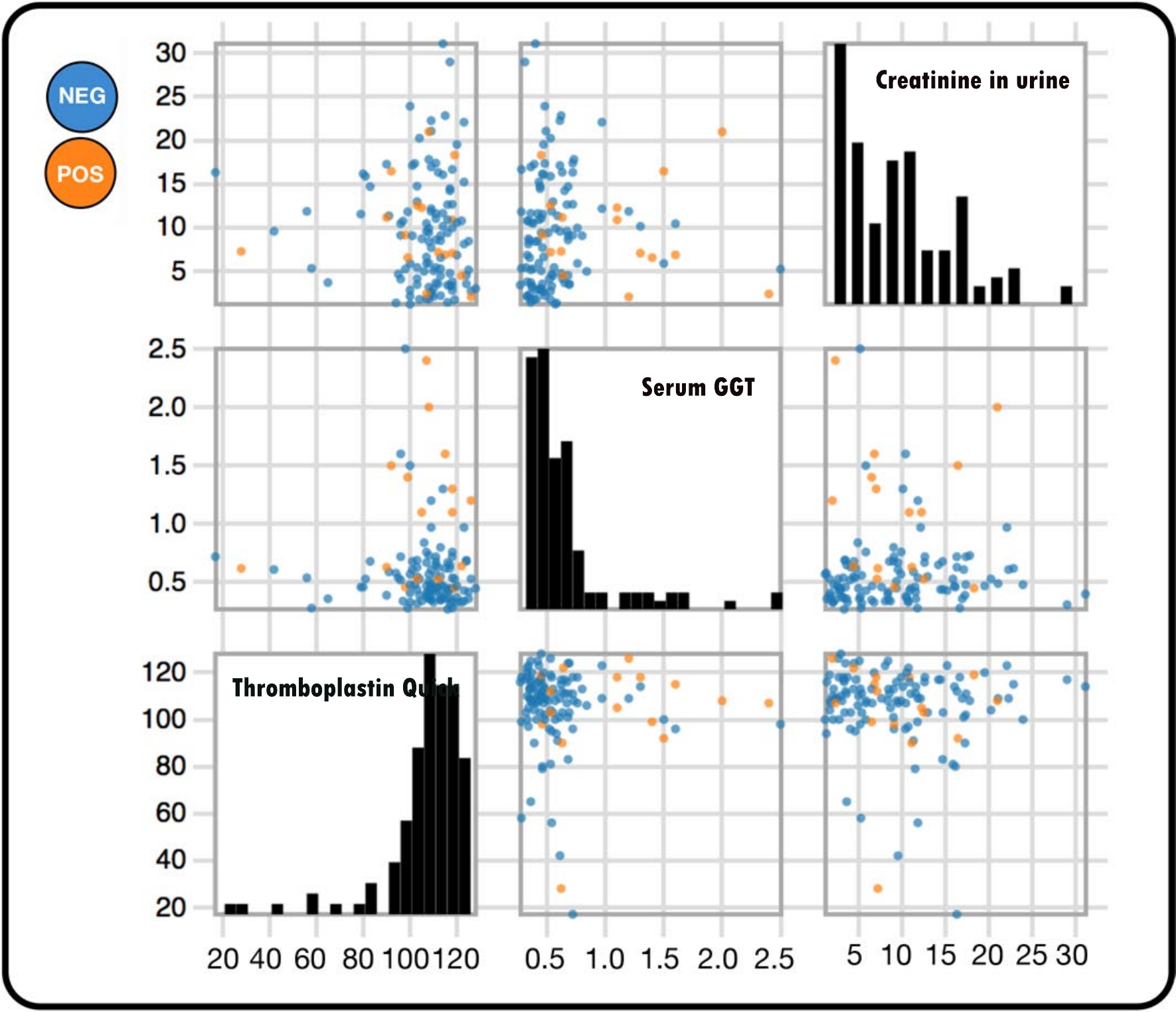}%
    \caption{\emph{\label{fig:scatter_s1} The distribution of S2-1 subspace cluster before transformation.}}}
\end{figure}

     \begin{figure}[h]
  \centering
  \includegraphics[width=1\columnwidth]{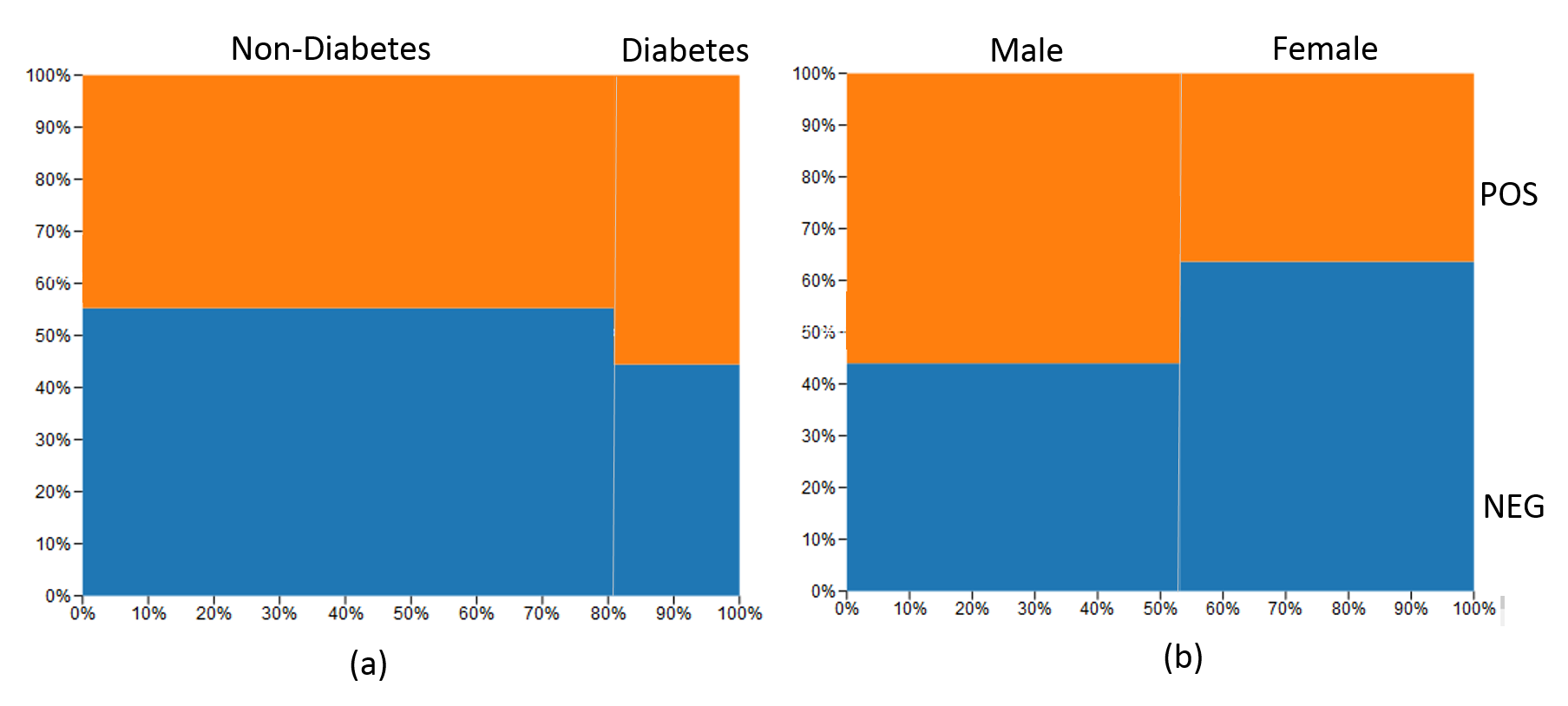}%
    \caption{\emph{\label{fig:mosaic}The proportion of participants regarding to diabetes and fatty liver (a) and the relationship between gender and fatty liver in the  oldest subspace population (b).}}
\end{figure}
\begin{figure*}[!h]
  \centering
  \includegraphics[width=\linewidth]{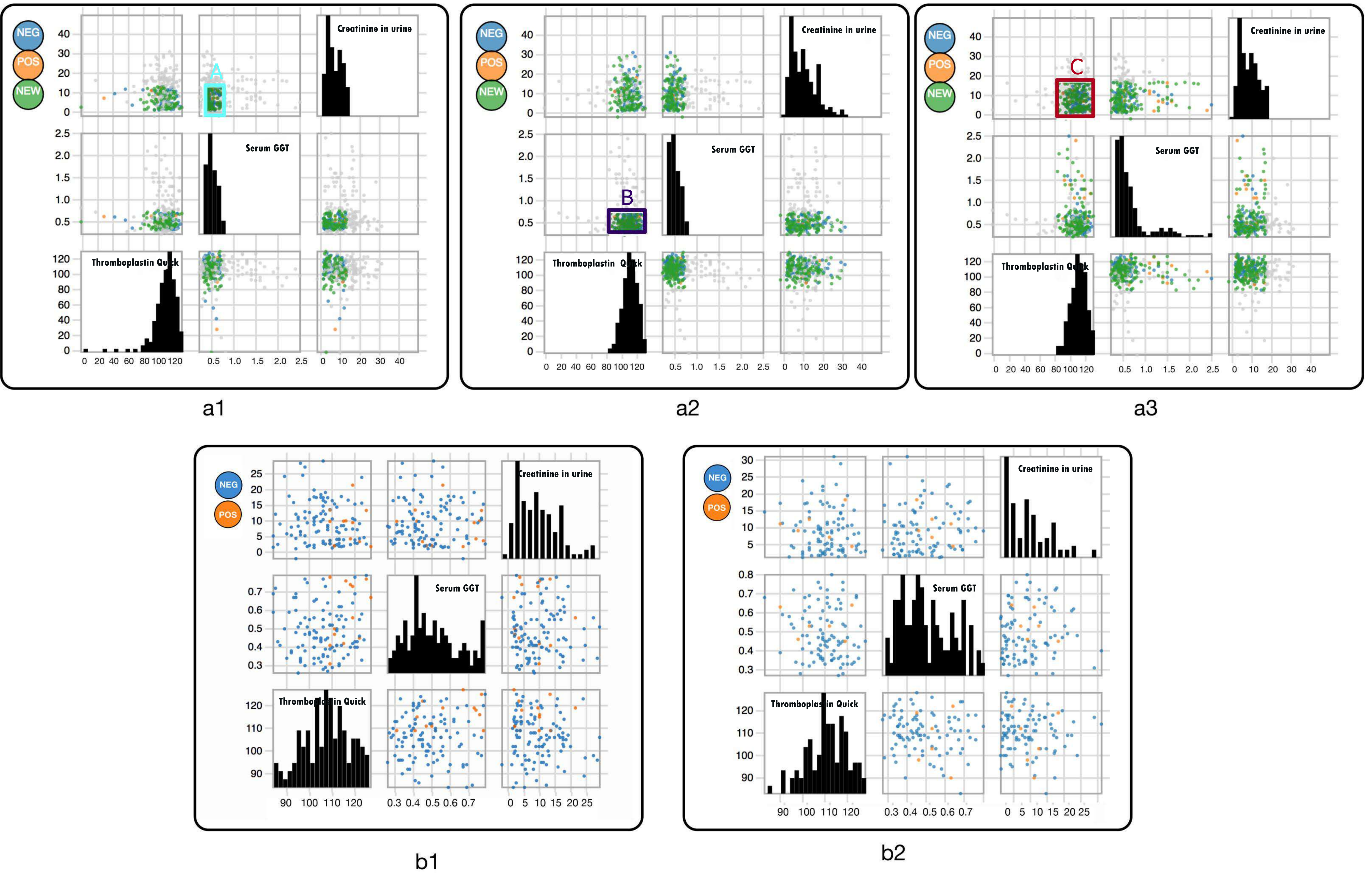}
  \parbox[t]{.9\columnwidth}{\relax
           }
           \caption{\label{fig:scattermatrix2}(a1)--(a3) illustrate the selected intervals. Fig.~(b1) is the distribution of the newly generated subpopulation labeled T-1. Fig.~(b2) illustrates the scatter matrix of the S2-1 subpopulation which is transformed to the new ranges.}
\end{figure*}
 \subsubsection{Exploration and Comparison of Subpopulations}
\begin{figure}[h]
  \centering
    \includegraphics[width=.8\columnwidth]{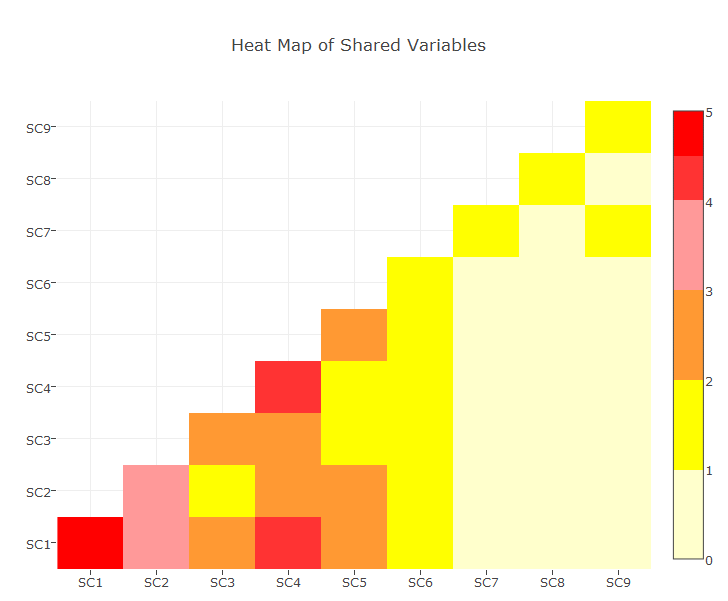}%
    \caption{\emph{\label{fig:SharedVariables}Heat Map of Shared Variables between Subspace Clusters}}
    \end{figure}

 As illustrated in Fig.~\ref{fig:screenshot}~(e), this part is used for analysis of subspace clusters and it is accessible via the chart panel.
 \begin{itemize}
     \item \emph{Donut heat map:} To support R\ref{depth}, we provided donut heat maps (Fig.~\ref{fig:screenshot}~(e)).
     In the donut heat map, sectors stand for variables. Rings represent individuals. Dimensions that do not contribute to one subspace cluster have gray scale coloring, and involved dimensions are mapped to colors. Darker colors depict greater values, in contrast, brighter ones stand for smaller values. 
     The user can zoom-in to see more detail and can select one participant (ring) to see its table of values in the footer section.
     
     There is also an optional sorting strategy based on the dimension that has the greatest variance. In this way, the participants that have greater value based on the dimension that has the greatest variance will have a bigger radius.
     \\
The following visualization techniques are employed to support R\ref{dimredundancy}-- R\ref{comparison}: 
     \item \emph{Mosaic Plot}: Mosaic charts are used to show the relationship of different nominal attributes. The user can dynamically select any two categorical variables (e.g. diabetes and fatty liver).       
    \item \emph{Heat map:} A heat map shows the object or dimension redundancies. Fig.~\ref{fig:SharedVariables} depicts dimension overlaps in subspace clustering result.
     \item \emph{Scatterplot matrix:} The distribution of numeric variables is shown via a discretized scatterplot matrix, equipped with histograms of each variable in the main diagonal. The separation of labels is provided by different colors. In Fig.~\ref{fig:scatter_s1}, discretization is based on the outcome fatty liver. The blue points are individuals with negative fatty liver and the orange ones show positive fatty livers. 
    \item \emph{Stacked errorbars:} Stacked errorbars compare different subspace clusters with the global mean based on any selected numeric variable in the whole data set. In Fig.~\ref{fig:ErrorBar}), each bar shows the proportion of positive and negative values of the outcome. Positive and negative values are normalized to form the average value of the corresponding cluster. All bars are sorted based on the average value and the error lines depict the standard deviation of the selected variable.
 \end{itemize}     

\section{Use Case Scenario}\label{sec:usecase}
To show how user can interact with S-ADVIsED to extract necessary information on subspace clustering results, we describe the system with two use case scenario: 

\begin{itemize}
\item Subpopulation discovery 
\item Exploration and comparison 
\end{itemize}

In all steps, expert users have an overall view of all subspace clusters (Fig.~\ref{fig:screenshot}~(d)). 
\subsection{Subpopulation discovery and validation} \label{sec:discovery}
In the first step, the analyst may want to select one interesting subspace cluster w.r.t. its similarity, size and involved dimensions and check its reproduciblity. By activating replication (Fig.~\ref{fig:screenshot}~(c)), the validation phase will be started and the analyst is able to specify the shape of cluster by selecting intervals and discover another subpopulation in another independent population. Thus, the analyst has to determine the number of candidate subpopulations. 
Next, the analyst selects a subspace cluster from the global view panel with considering its characteristics. We selected subspace cluster S2-1 (Fig. \ref{fig:screenshot}(d)). The distribution of S2-1 which is based on the results of DRESS is presented in Fig.~\ref{fig:scatter_s1}. Moreover, hypotheses about risk factors may focus on a specific group of participants, e.g. men aged over 50 years. Thus, in the second stage to transform the selected subspace cluster shape and validate it, we show the participants of S2-1 with the SHIP-TREND population which are located in S2-1 intervals in a combined view. To do this, a scatterplot matrix  defined by the variables of the selected subspace cluster with an embedded histogram is provided. Although, scatterplot matrices (SPLOMS) have limited scalability, as explained in Sec.~\ref{sec:requirements} we aim at low-dimensional subpopulations with up to 5 dimensions. Thus, for these data SPLOMS are appropriate. As presented in Fig. \ref{fig:scattermatrix2} (a1-a3) orange points are positive, and blue ones are negative fatty liver participants from SHIP-2 data. The green points are participants from SHIP-TREND. 
For the next step, the user can define the desired intervals for variables by drawing a rectangle in one pair of variables. Meanwhile, the user is drawing and expanding a rectangle she can see corresponding individuals which are located inside the rectangle w.r.t. the other pairs of involved variables via linking and brushing, see (Fig. \ref{fig:scattermatrix2} (a1-a3)). 

Next, the analyst is also wiling to find subpopulation in an independent cohort which is  similar to the original subpopulation regarding to the target variable (fatty liver) (recall Sec.~\ref{sec:validation}). Thus,
the labels of SHIP-TREND participants should be predicted based on the label of participants of the original S2-1 subpopulation which are located in the drawn rectangle. For this purpose, we assume that individuals closer to each other are more likely to have the same label. Thus, for label prediction we apply 1-nearest neighbor classification.

The user is enabled to draw multiple rectangles in different pairs of variables with distinct positions and diameters. Each drawn rectangle is a candidate to transform the selected subspace cluster and form a new subpopulation with SHIP-TREND participants within specified intervals. We have chosen three distinct rectangles in all pairwise variables. 

\begin{figure}[h]
  \centering
  \includegraphics[width=.7\columnwidth]{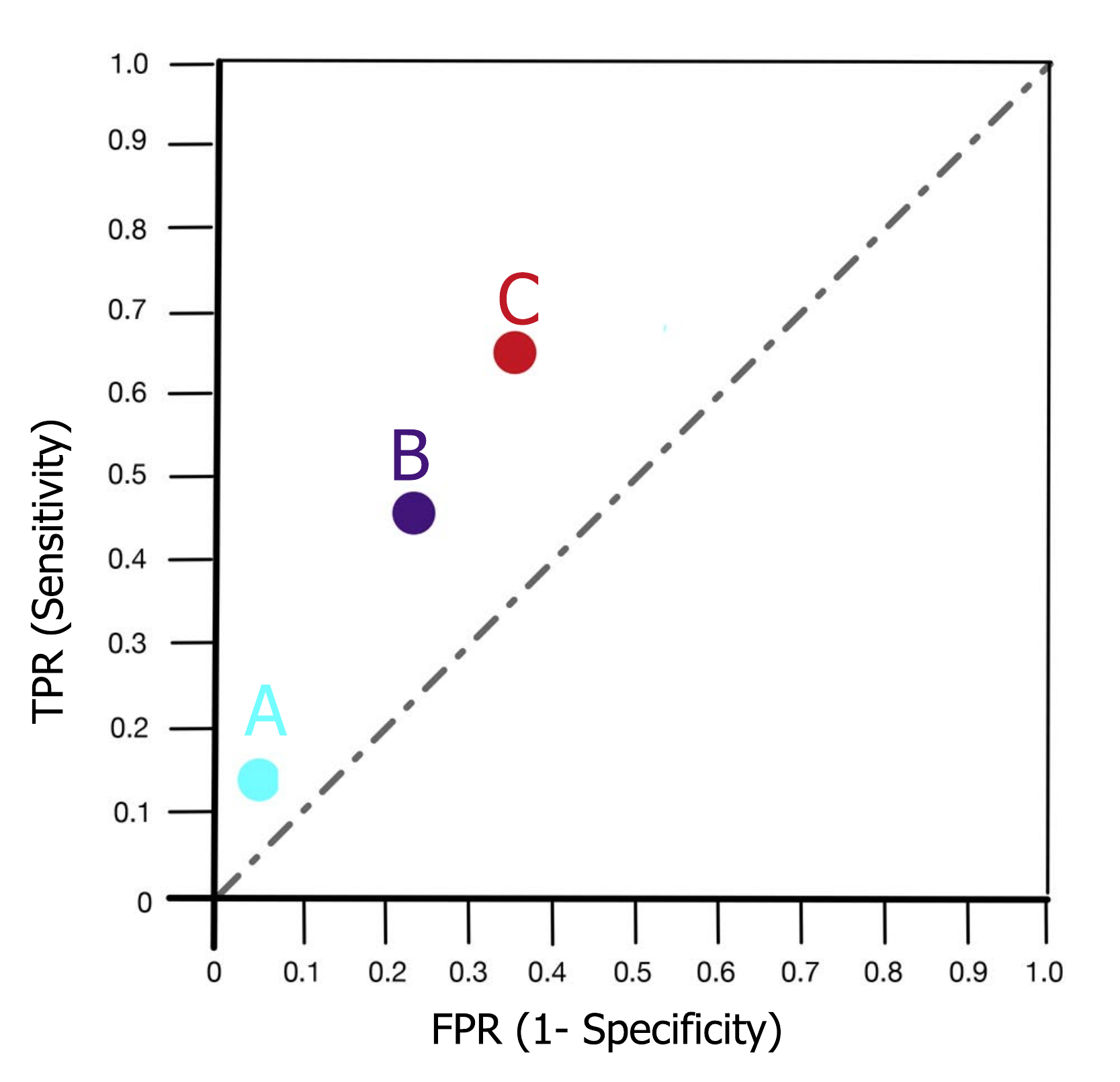}%
    \caption{\emph{\label{fig:roc}Each point in the ROC curve depicts the TPR and FPR ratio for the selected rectangle. Each point corresponds to a drawn rectangle in Fig.~\ref{fig:scattermatrix2} (a1-a3).}}
 \end{figure}

As next step, the epidemiologist should define which selected intervals are more appropriate to her in terms of boundaries and distribution regarding the outcome fatty liver. 
To do this, ROC curves show the relationship between true positive rate (tpr or sensitivity) and false positive rate (fpr or 1-specificity) (Fig.~\ref{fig:roc}). Then, the analyst selects one candidate subpopulation. Mostly, higher tpr and lower fpr are preferred because it means that both subpopulations have high percentage of fatty liver participants. 
In the next step, the new subpopulation is identified and the original subspace cluster will be transformed to the new intervals and form a subpopulation. In Fig.~\ref{fig:screenshot} (d), S2-1 is the transformed subpopulation and T-1 is the new subpopulation.
Finally, both the transformed original subpopulation(S2-1) and new discovered subpopulation(T-1) will be shown in the global view for further investigations. So that, the analyst can check whether the subpopulation description is the same in both subpopulations (recall Sec.~\ref{sec:validation}). 

\subsection{Exploration and Comparison}
\begin{figure*}[!h]
     \centering
     \includegraphics[width=.9\columnwidth] {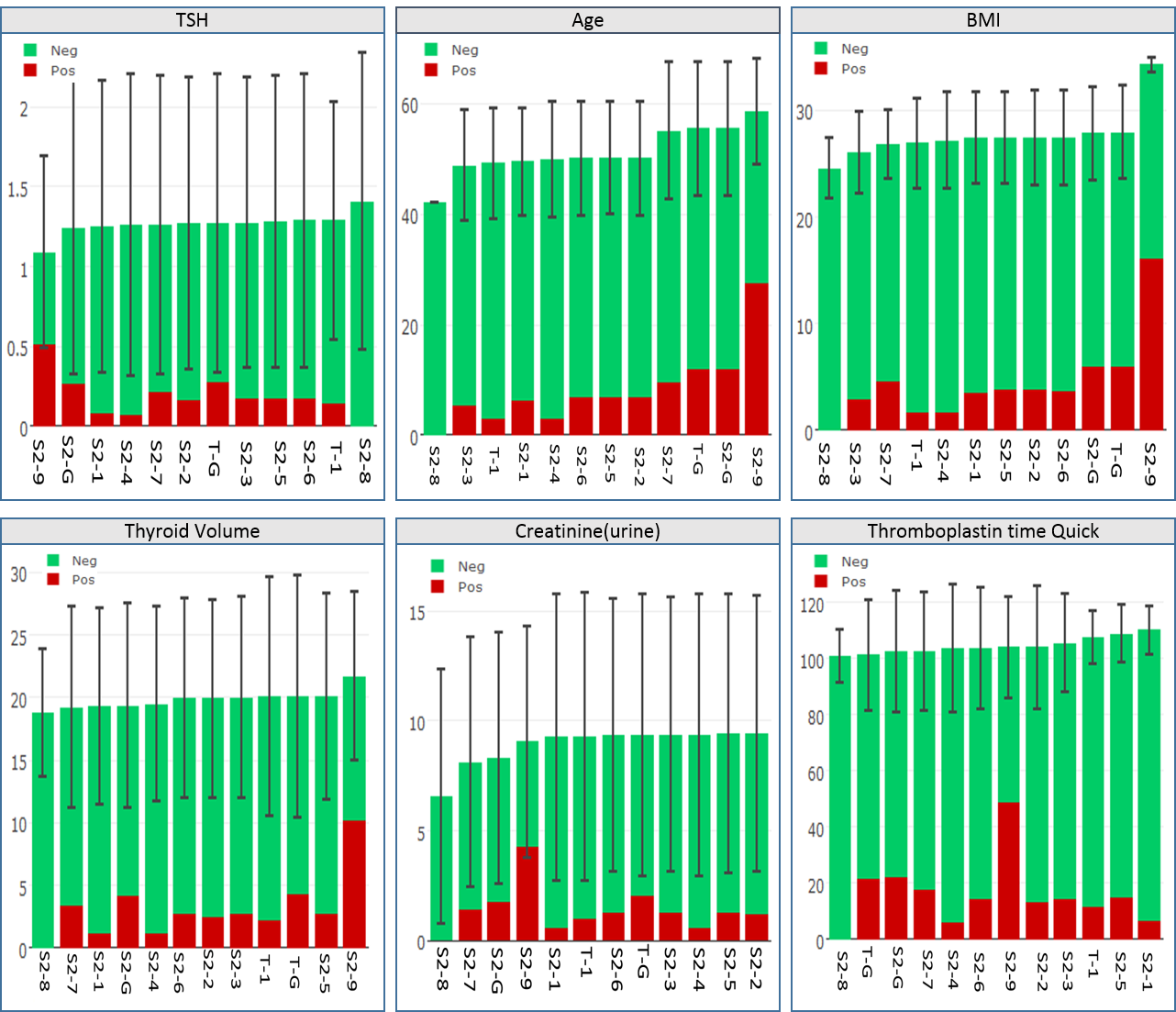}  
      \parbox[t]{.5\columnwidth}{\relax
           }
           \caption{\label{fig:ErrorBar} The stacked error bar shows the mean value of different variables for all subspace clusters. The error line depicts the standard deviations for selected variable.
           }
    \end{figure*}

The analyst can inspect the characteristics of subspace clusters and also check the reproducibility of subpopulations gained in the replication phase.
In other words, each subspace cluster has a cluster description and after subpopulation discovery and replication phase the analyst should check that the original and new subpopulation in the independent cohort have the same subpopulation description.
To start, the analyst may be interested to see the similarity between subspace clusters separately with respect to the shared variables or participants. 
Apart from the global view which shows the similarity based on a trade-off between shared variables and participants using the slider, she can select the heat map from the charts panel (Fig.~\ref{fig:screenshot}~(b)(1)).
As next step, by zoom-in and tooltips (to see involved variables) she selects a subspace cluster based on the interestingness regarding its involved variables, number of participants and similarity to the other subspace clusters. For example, we selected S2-1 with three numeric (Thromboplastin time Quick, serum GGT, creatinine in urine) and two nominal (smoking status, enalapril) variables, where all participants are ex-smokers.\\
Next, by selecting the donut heat map from the charts panel the user can get an overview of the selected subspace cluster in a compact view via an in-depth analysis panel. The analyst can interactively click on a participant (ring) and see the table of values in the footer. Moreover, all subspace clusters that share this participant will be highlighted in the global view panel (Fig.~\ref{fig:screenshot}~(d)).\\
For the next step, the analyst wants to compare subspace clusters with respect to different variables. When variables are categorical, the analyst selects the mosaic plot from the charts panel and then the other variable from the submenu. Then, she can see the relationship between target variable and selected variable (i.e. the percentage of female and male participants w.r.t. positive and negative fatty liver) (Fig.~\ref{fig:screenshot}~(b)(4)). If the analyst wants to see the distribution of involved numeric variables of a subspace cluster, then she selects distributions and then a categorical variable as discretization parameter for coloring the points in scatter the plot matrix (Fig.~\ref{fig:screenshot}~(e)(3)). In this way, the analyst can also compare the subpopulations for replication (i.e. is the distribution the same in both subpopulations regarding to the target variable?. 
\\
As the next step, the analyst may want to compare all subspace clusters w.r.t. the mean value of a specific measurement and their distance to the global mean. To do this, she selects a stack error bar and selects a variable from submenu for comparison (Fig.~\ref{fig:screenshot}~(e)(5)). Additionally, as mentioned in Sec.~\ref{sec:validation} the mean values of replicated and the original subpopulation should deviate in a similar way from the global mean values. As shown in Fig.~\ref{fig:ErrorBar} the analyst can check this factor regarding to different variables using errorbars.

\section{Results}
\label{sec:results}
Instead of an evaluation, we describe results of the subspace clustering achieved by the DRESS algorithm using our S-ADVIsED framework. The final decision for replication of a subpopulation needs expert knowledge in epidemiology.

To analyze the subspace clusters, we started with the validation phase. 
As mentioned in Sec.~\ref{sec:discovery} to show how S-ADVIsED works, we have chosen cluster S2-1 (Fig.~\ref{fig:screenshot}). It has three numeric (Thromboplastin time Quick, serum GGT, creatinine in urine) and two nominal (smoking status, enalapril) variables.

Since our approach for replication is based on the projection of numeric variables, we filtered the set of SHIP-TREND participants regarding to the nominal variables of the selected subspace cluster. For example, all participants in the selected cluster are ex-smokers, so we just considered all participants that are ex-smokers in the projection of the scatterplot matrix. %
The distribution of S2-1 is shown in Fig.~\ref{fig:scatter_s1}. We have dense and sparse parts in all pairs of involved variables. Fig.~\ref{fig:scattermatrix2} (a1--a3) indicates the union of participants in S2-1 and SHIP-TREND. To transform the boundaries for S2-1 and identify a new subpopulation, we have chosen the subpopulation (rectangle) that in comparison with other candidate subpopulations, is far from a diagonal line in ROC (higher sensitivity). \\
We compared the two subpopulation regarding different measurements (recall Sec.~\ref{sec:validation}) 
Fig.~\ref{fig:scattermatrix2} (b1-b2) shows the distribution of the transformed selected subpopulation (S2-1) and the newly generated subpopulation (T-1), respectively. Both subpoulations have apparently equivalent distributions in all pairs of variables and regarding to the percentage of negative/positive fatty liver participants.
Regarding to the dimentionality, as only involved variables of S2-1 are consider in T-1, the number of variables in both subpopulations is the same.
Moreover, after the subpopulation description phase S2-1 and T-1 have 95 and 104 participants, respectively. %
From the deviation perspective, we compared the whole SHIP-2 and SHIP-TREND with S2-1 and T-1 subpopulations regarding to invloved variables. The average levels of serum GGT in the whole SHIP-2 (S2-G) and SHIP-TREND (T-G) populations are 0.72 and 0.65 $\mu mol/sl$, respectively, whereas it is 0.48 $\mu mol/sl$ in subpopulation S2-1 and 0.49 $\mu mol/sl$ in T-1.  Fig.~\ref{fig:ErrorBar} shown that the mean value of creatinine in urine is 8.3, 9.35, 9.25 and 9.29 $mmol/l$ in S2-G, T-G, S2-1 and T-2, respectively. The subpopulations S2-1 and T-1 have nearly similar values w.r.t. thromboplastin time quick and deviate the same way from S2-G and T-G.
     
Additionally, we also analyze subpopulation S2-9 which has the greatest proportion of  positive fatty liver (Fig.~\ref{fig:ErrorBar}). 6.7 percent of SHIP-2 individuals belong to S2-9. Additionally, more than 50$\%$ of male participants suffer from fatty liver disorder and about 18$\%$ are diagnosed as diabetic while in the whole SHIP-2 population the portion of diabetes is 10$\%$ Fig.~\ref{fig:mosaic}.

As illustrated in Fig.~\ref{fig:ErrorBar}, S2-9 is the oldest subpopulation with an average age of 58$\pm$9 years. All participants in this subpopulation suffer from obesity with a mean BMI of 34.2$\pm$0.06 $kg/m^2$. They also deviate strongly from the global mean regarding the thyroid-stimulating hormone (TSH) by a mean value of 1.09$\pm$0.6 $mU/l$. It has the lowest average TSH value and on average a bigger total thyroid volume with 21.7$\pm$6.7 $Ng/dl$.

\section{Conclusion \& Future Work}
\label{sec:conclusion}
In this work, we presented S-ADVIsED as a web-based visualization framework for exploration of subpopulations. The design of the system was based on site visits at the epidemiology department and is largely based on ideas of epidemiologists, e.g. for transforming clustering results in subpopulations and in validating such subpopulations. 
 %

To extract a comprehensible description of an arbitrarily shaped subspace cluster, we have included a mechanism letting epidemiologists interactively approximate the cluster's linear boundaries by drawing rectangles w.r.t. the involved variables (recall Fig.~\ref{fig:roc}). %
We intend to develop a method that maximizes the product of sensitivity and specificity delivering an recommended hyper-rectangular approximation of a subpopulation which subsequently can be adjusted based on the epidemiologist's suggestion. 

For future work, we intend to apply a post-filtering step to finalize the set of subspace clusters: We will arrange groups of subspace clusters with large participants and variable overlaps in a hierarchical way, such that the expert can expand individual subspace clusters of interesting groups on demand. Furthermore, we will develop density plots instead of scatterplots to better support the comparison of the two distributions.
\section*{Acknowledgement}\label{sec:Acknowledgement}
This paper is an extension of work originally reported in \emph{"EuroVis Workshop on Visual Analytics"}~\cite{AlemzadehEtAl:EuroVA2017}.\\
SHIP is part of the Community Medicine Research net of the University of Greifswald, Germany, which is funded by the Federal Ministry of Education and Research (grant no.
03ZIK012), the Ministry of Cultural Affairs as well as the Social Ministry of the Federal State of Mecklenburg-West Pomerania.
\bibliographystyle{eg-alpha-doi}
\bibliography{egbibsample}
\end{document}